\documentclass[a4paper]{article}

\usepackage[english]{babel}
\usepackage{icrc2013}

\title{Novel Photo Multiplier Tubes for the Cherenkov Telescope Array Project}

\shorttitle{ICRC 2013 Template}

\authors{
takeshi toyama$^{1}$,
razmik mirzoyan$^{1}$,
hugh dickinson$^{1,2}$,
christian fruck$^{1}$,
j\"{u}rgen hose$^{1}$,
hanna kellermann$^{1}$,
max kn\"{o}tig$^{1}$,
eckart lorenz$^{1}$,
uta menzel$^{1}$,
daisuke nakajima$^{1}$,
reiko orito$^{3}$,
david paneque$^{1}$,
thomas schweizer$^{1}$,
masahiro teshima$^{1,4}$,
tokonatsu yamamoto$^{5}$,
for the CTA Consortium
}

\afiliations{
$^1$ Max-Planck-Institut for Physics \\
$^2$ The Oskar Klein center for cosmoparticle physics \\
$^3$ Faculty of Integrated Arts and Sciences, The University of Tokushima \\
$^4$ Institute for Cosmic Ray Research, University Tokyo \\
$^5$ Department of Physics, Konan University \\
}

\email{taktoy@mpp.mpg.de, razmik@mpp.mpg.de}

\abstract{Currently the standard light sensors for imaging atmospheric Cherenkov telescopes are the classical photo multiplier tubes that are using bialkali photo cathodes. About eight years ago we initiated an improvement program with the Photo Multiplier Tube (PMT) manufacturers Hamamatsu (Japan), Electron Tubes Enterprises (England) and Photonis (France) for the needs of imaging atmospheric Cherenkov telescopes. As a result, after about 40 years of “stagnation” of the peak Quantum Efficiency (QE) on the level of 25-27\%, new PMTs appeared with a peak QE of 35\%. These have got the name “super-bialkali”. The second significant upgrade has happened very recently, as a result of a dedicated improvement program for the candidate PMT for Cherenkov Telescope Array. The latter is going to be the next generation major instrument in the field of very high energy gamma astrophysics and will consist of over 100 telescopes of three different sizes of 23m, 12m and 4-7m, located both in southern and northern hemispheres. Now PMTs with average peak QE of approximately 40\% became available. Also, the photo electron collection efficiency of the previous generation PMTs of 80- 90\% has been enhanced towards 95-98\% for the new ones. The after-pulsing of novel PMTs has been reduced towards the level of 0.02\% for the set threshold of 4 photo electrons. We will report on the PMT development work by the companies Electron Tubes Enterprises and Hamamatsu Photonics K.K. show the achieved results and the current status.
}


\begin{document}
\maketitle

\section{Introduction}
Cherenkov Telescope Array (CTA) \cite{bib:CTA} is the next generation Imagining Atmospheric Cherenkov Telescope (IACT) project for achieving 10 times better sensitivity than the current major IACTs (H.E.S.S. \cite{bib:hess}, MAGIC \cite{bib:magic} and VERITAS \cite{bib:veritas}). For successful realization of this project, many hardware improvements were necessary.

We started a special development program with the PMT manufacturers Hamamatsu Photonics K.K. (Japan) and Electron Tubes Enterprises Ltd. (England) for novel PMTs for the CTA project. We believed one could improve all the major parameters of PMTs and that will boost the sensitivity and widen the covered energy range of the CTA telescopes. We have defined and requested from the above companies an extensive program for improving multiple parameters of PMTs. Some of the main requested parameters are listed in Table\ref{table_FPI_spec} below.

\begin{table}[!h]
\begin{center}
\begin{tabular}{|l|c|c|}
\hline Parameter & Specification \\ \hline
Spectral Sensitivity Range  & 290 - 600 nm  \\ \hline
Peak Quantum Efficiency   & $\geq$35\%  \\ \hline
Average QE over Cherenkov Spectrum &  $\geq$21\%  \\ \hline
Afterpulsing at 4 ph.e. Threshold  &  $\leq$0.02\%  \\ \hline
Transit Time Spread, single ph.e, FWHM  &  $\leq$1.5 ns  \\ \hline
Collection Efficiency 1.st Dynode  & $\geq$95\%  \\ \hline
\end{tabular}
\caption{The requested specifications for the CTA PMTs}
\label{table_FPI_spec}
\end{center}
\end{table}

As shown in Table \ref{table_FPI_spec}, along with the Photo Detection Efficiency(PDE), which includes the Quantum Efficiency (QE) and the Collection efficiency, we requested to significantly reduce the after pulse rate; only under that condition can one really profit from the enhanced PDE. Please note that the effect of afterpulsing is a major parameter for defining the threshold energy of an IACT [5].  

\section{Quantum Efficiency Measurement of Photo Multiplier Tubes}
\begin{figure}[th]
  \centering
  \includegraphics[width=0.5\textwidth]{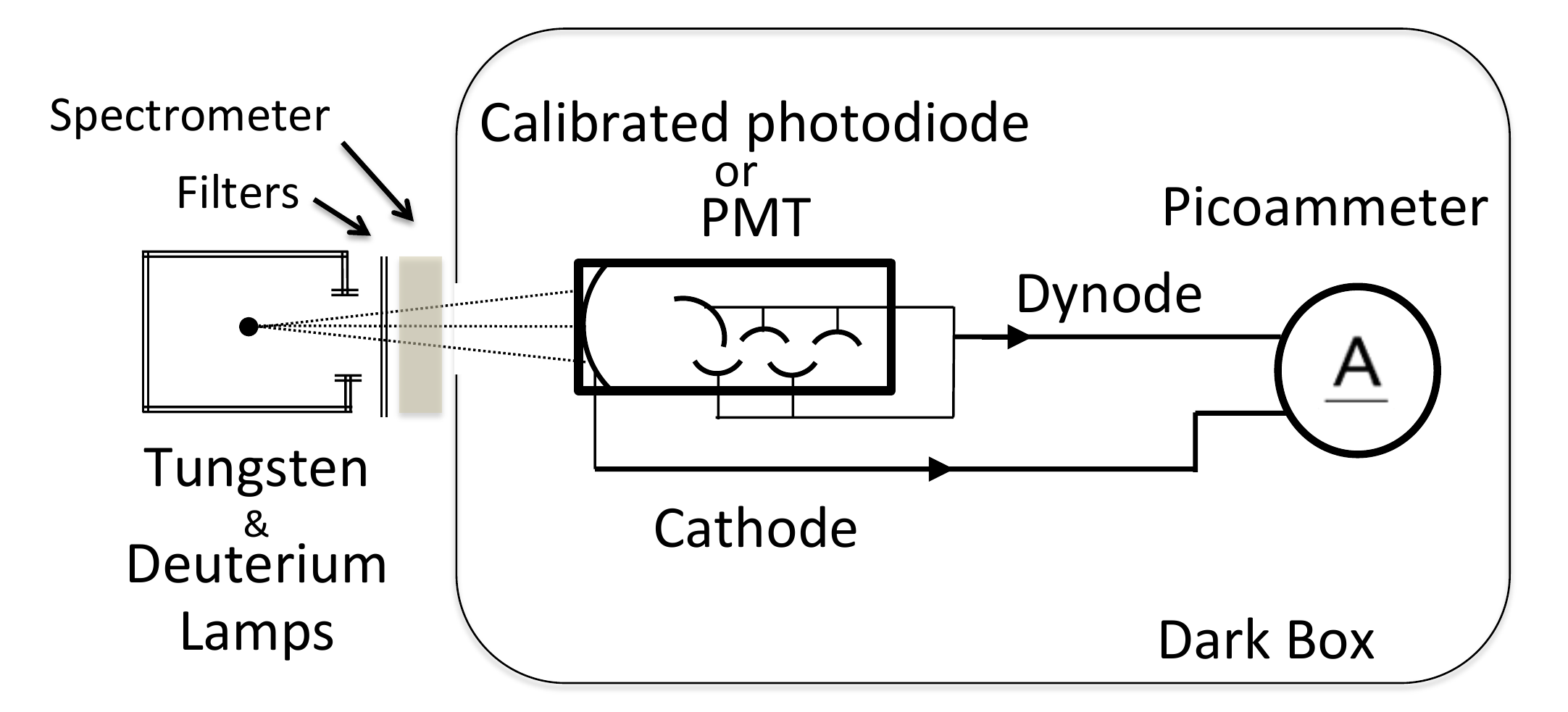}
  \caption{Schematics of the QE measuring device}
  \label{fig_exp_setup_QE}
 \end{figure}

We have constructed a Quantum Efficiency measurement device(Figure \ref{fig_exp_setup_QE}) that consists of a) a light source box, hosting a Tungsten and a Deuterium lamps, b) a modified commercial spectrometer with three different gratings, c) a rotating filter wheel for suppressing the unwanted wavelengths produced by the gratings and d) a metallic dark box enclosing the light sensor under test and e) a calibrated calibrated PIN diode of a tabulated QE for every 10nm.

While illuminating the test sensor and the calibrated diode with all the wavelengths in the range of interest we measure their output current by using Keitley picoammeters of type 6485. In the measurement of the QE of a selected PMT we measure the current flowing between the cathode and the first dynode; rest of the dynodes are shorted with the first one for avoiding space charge effects that can influence our measurements. The actual QE of a PMT is calculated by comparing its photo cathode current with that of a reference calibrated PIN photo diode.

\begin{figure}[h]
  \includegraphics[width=0.5\textwidth]{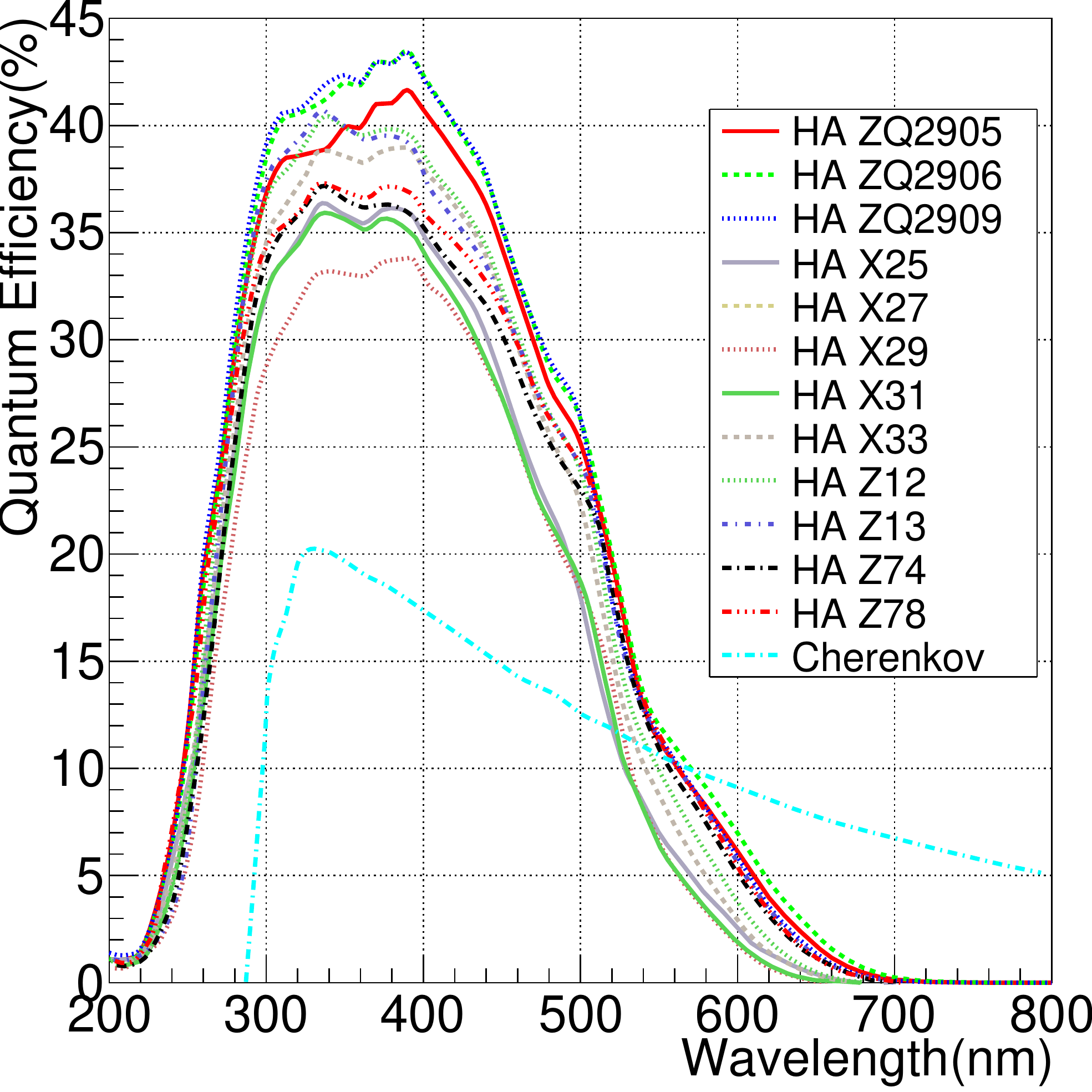}
  \caption{Quantum Efficiency of 3 Hamamatsu R11920-100-05 (ZQ2905, ZQ2906, ZQ2909) and 9 Hamamatsu R11920-100 PMTs measured over wavelength 200 - 800nm. The bottom dashed line is the simulated Cherenkov light spectrum of 100GeV air showers from zenith measured at 2km a.s.l..}
  \label{fig_QE_hamamatsu}
 \end{figure}

We have measured 9 samples of Hamamatsu PMTs (R11920-100) and 3 samples of new Hamamatsu PMTs (R11920-100-05) that showed peak QEs over peak QE over 35\%, see Figure \ref{fig_QE_hamamatsu}. Serial numbers of new PMTs are shown as ZQ29XX. New PMTs ZQ2906 \& ZQ2909 show a QE in excess of 43\%. The bottom dashed line shows the spectrum of Cherenkov light from 100GeV air showers coming from near zenith direction. The simulated altitude is 2km a.s.l.. As one can see, the PMT QE curves make a good match to the simulated Cherenkov light spectrum. 

\begin{figure}[th]
  \includegraphics[width=0.48\textwidth]{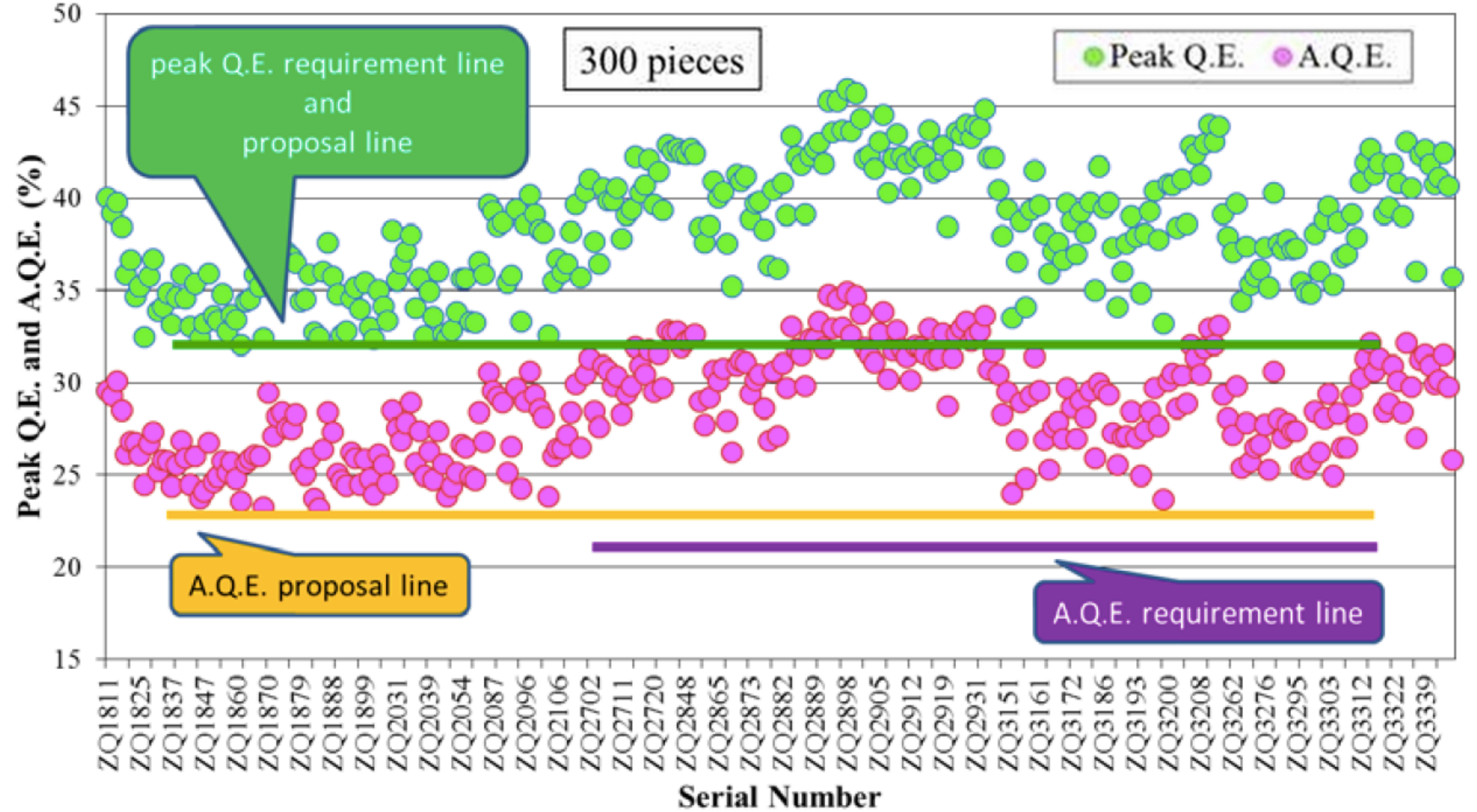}
  \caption{Quantum Efficiency of 300 recent PMTs (R11920-100-05) from Hamamatsu. green circles: peak QE. magenta dot: Average QE over Cherenkov spectrum (290nm-600nm). The purple line shows the average over Cherenkov spectrum QE request by CTA. The green and the yellow lines show correspondingly the minimum peak and the average over the Cherenkov spectrum QEs offered by Hamamatsu.}
  \label{fig_Hama300piecesPMTsQE}
 \end{figure}

Hamamatsu provided us with a figure showing the QEs of recently produced 300 PMTs, see Figure \ref{fig_Hama300piecesPMTsQE}. One can see that there are quite some number of PMTs showing peak QEs in excess of 40\%. The company could meet and even further improve our requirements on QE and also on folded with Cherenkov spectrum QE, see the green and yellow lines on \ref{fig_Hama300piecesPMTsQE}. 
 
\begin{figure}[!h]
  \includegraphics[width=0.5\textwidth]{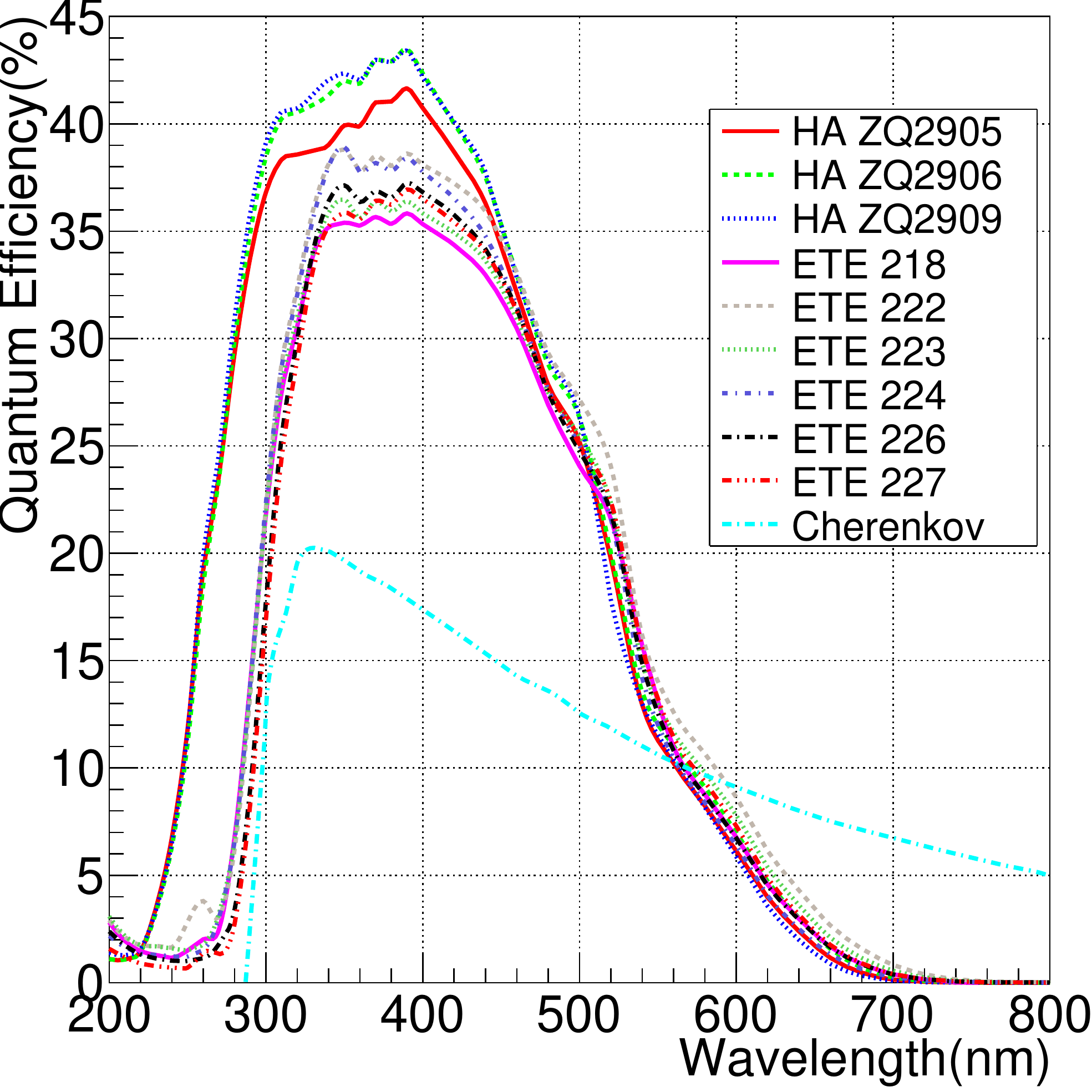}
  \caption{The QE of 6 PMTs of type D872/2A from ETE and 3 from Hamamatsu  (ZQ2905, ZQ2906, ZQ2909) for the wavelength range of 200-800nm is shown. The bottom dashed line is the simulated Cherenkov light spectrum of 100GeV air showers from zenith measured at 2km a.s.l..}
  \label{fig_QE_electrontube}
 \end{figure}

We measured also 6 samples of Electron Tubes Enterprises(ETE)  PMTs (D872/2A) (Figure \ref{fig_QE_electrontube}). The peak QE of these tubes are in the range of 35-37\% in the wavelength range 350-400nm. These values are somewhat lower than those from Hamamatsu PMTs. For further enhancing the QE ETE is planning to use anti-reflective coat between the photo cathode and the front glass. This should increase the QE by 10-12\%. Also, ETE is planning to use PMTs of mat input window. This should further enhance the QE by another 6-8\%. Thus we expect that in the end of the development work both types of PMTs will have similar QEs. The relatively low QE of ETE PMTs at the 300nm and below is due to the used glass type.

 \begin{figure*}[]
  \centering
  \includegraphics[width=0.67\textwidth]{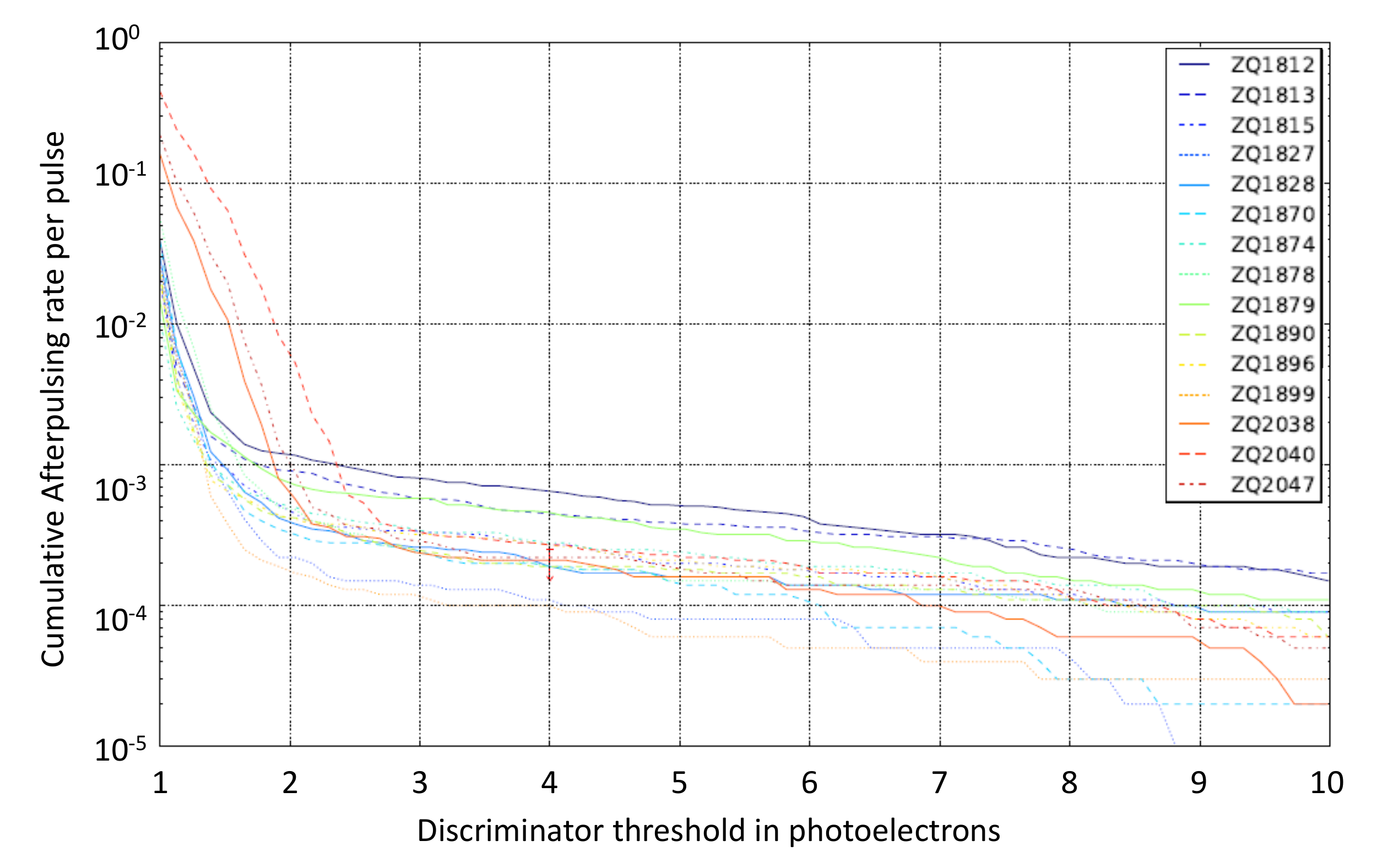}
  \caption{Afterpulse rate versus photoelectron threshold level.}
  \label{fig_afterpulse}
 \end{figure*}

\section{Afterpulse of Photo Multiplier Tubes}

In Figure \ref{fig_afterpulse} we show the measured afterpulse rate versus the discriminator threshold (in photo electrons). The afterpulse events are searched for the duration of up to 2$\mu$s after the impinging laser shots. Although, the full probability spectra of alterpulses as a function of its amplitude should be taken into account, we can use the rate above the threshold of 4 photo electrons (ph.e.) as a figure of merit to characterize the PMTs. The requested in Table \ref{table_FPI_spec} afterpulse rate of 0.02\% (or $2 \times 10^{-4}$ in Figure \ref{fig_afterpulse}) is normalized to the rate of induced single ph.e.s. The main reason of afterpulses in PMTs, in spite of deep vacuum, is due to the ionization of residual atoms and molecules that are hit by accelerated electrons. These atoms and molecules exist inside the envelope of PMT in a gas form or are adsorbed in the material of dynodes. These positive ions ($\rm{H^{+}}$, $\rm{He^{+}}$, $\rm{CH_{4}^{+}}$, e.t.c., ) that are much heavier than the electrons, are accelerated and hit the cathode releasing bunches of multiple electrons.  

 \begin{figure}[!h]
  \centering
  \includegraphics[width=0.48\textwidth]{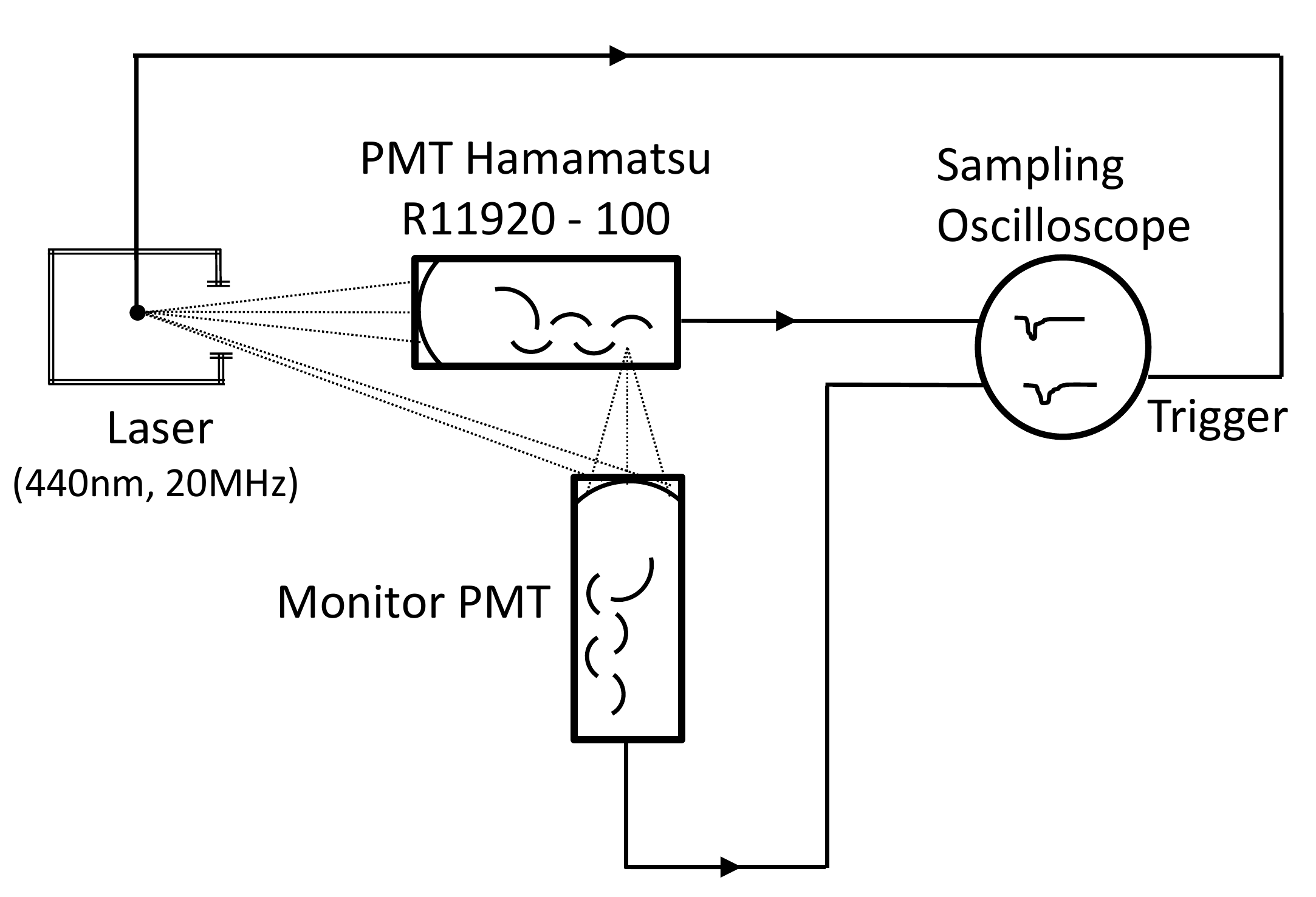}
 \caption{Schematics of the set-up to measure dynode glow} 
  \label{fig_exp_setup_light_emission}
 \end{figure}

Hamamatus tried hard to reduce the afterpulses in CTA candidate PMTs. For this purpose they have used a dynode structure from another PMT, namely R 8619, that was known for its very low afterpulse rate. Along with that they started using 4 large getters and a stronger evacuation of PMT glass bulbs. Still the afterpulse rate was not as low as requested by us. It became necessary to further reduce the afterpulse rate by alternative ways. Another possible reason of afterpulses is due to the accelerated in inter-dynode space electrons that hit the sequential dynodes and can produce light, presumable via bremsstrahlung. Part of this light can find its way back to the photo cathode (the system of dynodes can guide the light back to the photo cathode) and kick out new electrons.

 \begin{figure}[h]
  \centering
  \includegraphics[width=0.48\textwidth]{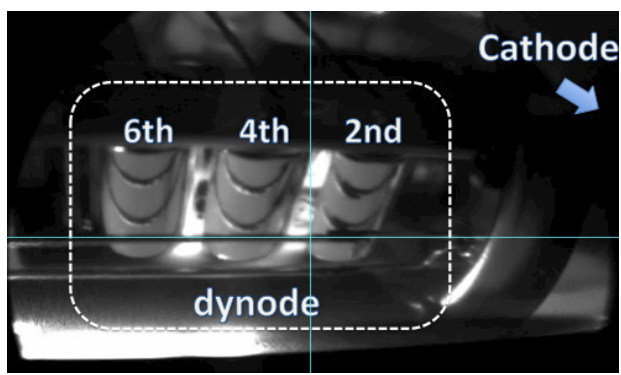}
  \caption{This photo documents the light emission from inter-dynode space. Laser pulses impinge from right, and light emission can be seen between the dynodes 2 \& 4 and 4 \& 6. }
  \label{fig_dynode_emisson}
 \end{figure}

 \begin{figure}[h!]
  \centering
  \includegraphics[width=0.48\textwidth]{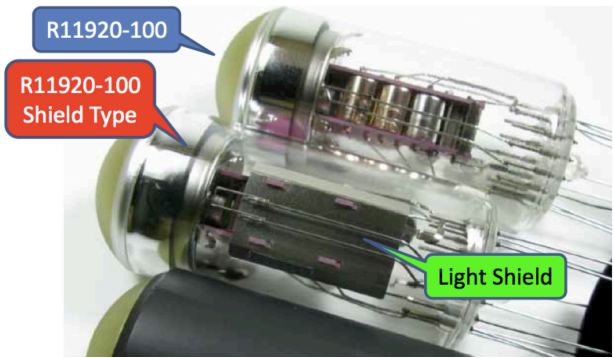}
  \caption{Comparison of Hamamatsu R11920-100 (top) with R11920-100 of Shield Type (bottom) The shield reduces the light-induced after-pulsing.}
  \label{fig_PMT_pic}
 \end{figure}

 To check a light emission from inside PMT, an experiment is set up (Figure \ref{fig_exp_setup_light_emission}). There are 2 PMTs in this setup, one, Hamamastu PMT (R11920-100), is facing the laser, while the other ,monitor PMT, is facing the dynodes of the first PMT to detect a light emission from the dynodes. A laser operating at 440nm wavelength is shooting to a photo cathode of the first PMT and a part of its radiation can reach to photo cathode of Monitor PMT. Both signals of PMTs are fed into an oscilloscope which is triggered from the laser signal. In monitor PMT output, we can distinguish "a signal by laser" from "a signal by a emission from the dynodes" from timing information. On Figure \ref{fig_dynode_emisson}, we show a photo of light emission from between the dynodes of a Hamamatsu PMT. The latter is illuminated with a laser light of 440nm at a rate of 40MHz, the exposure of the photo is 1s. The laser light is impinging from the right side. One can see clear light emission emerging between the dynodes 2nd \& 4th and 4th \& 6th.

 \begin{figure}[h]
  \centering
  \includegraphics[width=0.48\textwidth]{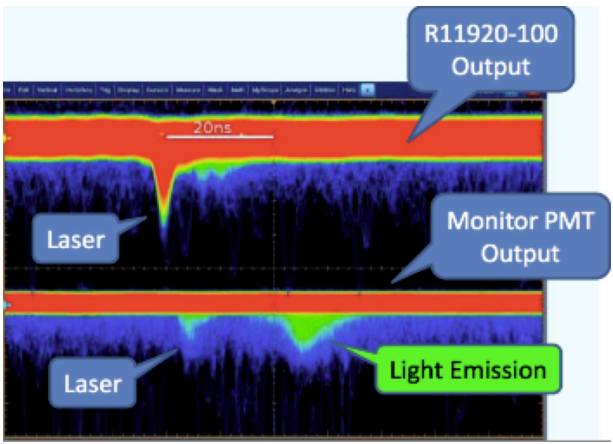}
  \caption{upper: the first PMT measures pulses from laser in persistence mode. lower: the monitor PMT measure light emission from dynodes of the first PMT. After 18ns the dynodes start emitting light.}
  \label{fig_afterpulse_signal1}
 \end{figure}
\begin{figure}[h!]
  \centering
  \includegraphics[width=0.5\textwidth]{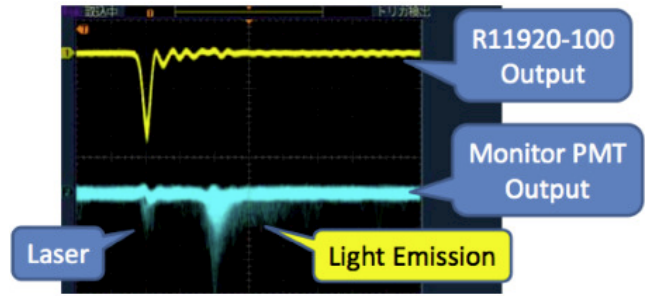}
  \caption{top:light signal measured by a Hamamatsu PMT (R11920-100). bottom: Signal from a monitor PMT}
  \label{fig_PMT_afterpluser}
 \end{figure}
\begin{figure}[h!]
 \flushright
  \includegraphics[width=0.45\textwidth]{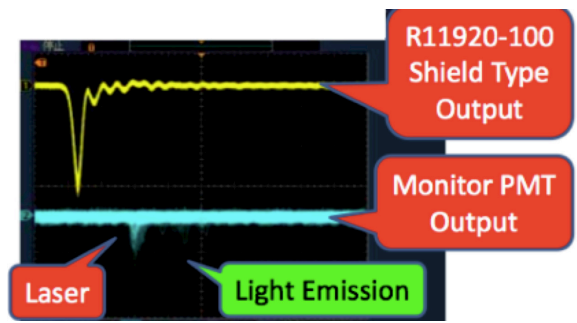}
  \caption{top:Signal from a Hamamatsu PMT (R11920-100) that uses a shield. bottom: Signal from the monitor PMT.}
  \label{fig_PMT_afterpluser_improved}
 \end{figure}

Hamamatsu tried to reduce this problem by shielding the dynodes. On Figure \ref{fig_PMT_pic} one can see the dynodes of the "old" version of PMT (the upper one) while they are covered by a shield for the new generation of PMTs (the lower PMT). Hamamatsu repeated the measurements of MPI for Physics. Their measurements have confirmed our findings, see Figure\ref{fig_PMT_afterpluser} . The Light Emission is clearly reduced as a result of using a new shield for the PMT R11920-100 ( Figure \ref{fig_PMT_afterpluser_improved} ). To check this in more detail, a few PMTs are shipped to our institute. 

In Figure \ref{fig_afterpulse_signal1} , the oscilloscope output in persistence mode shows the signals for both the first PMT on the upper trace and the monitor PMT on the lower trace. The vertical and horizontal axes show voltage and time. The monitor PMT detects a light emission from dynodes after $\sim$ 18nsec.

\

\section{Conclusions}
We have evaluated PMTs for the next generation Very High Energy gamma ray observatory, CTA. QE of both PMTs by Hamamatsu Photonics K.K. and Electron Tubes Enterprises Ltd are over 35\% at peak. We have also measured afterpulse rate of PMT  R11900-100 by Hamamatsu, As the result afterpulse rate is higher than required specification by CTA project. Light emission on dynodes is found to be cause of afterpulse. Hamamatsu have improved afterpulse rate of R11900-100 by setting a shield around dynodes. 



\vspace*{0.5cm}
\footnotesize{{\bf Acknowledgment:}{We gratefully acknowledge support from the agencies and organizations listed in this page: http://www.cta-observatory.org/?q=node/22}


\begin{thebibliography}{}
\bibitem{bib:CTA} The CTA Consortium. arXiv:1008.3703
\bibitem{bib:hess}”The H.E.S.S. project”.http://www.mpi-hd.mpg.de/hfm/HESS/
\bibitem{bib:magic}”The MAGIC Telescope”.http://wwwmagic.mppmu.mpg.de/
\bibitem{bib:veritas}”The VERITAS project”.http://veritas.sao.arizona.edu/
\bibitem{bib:NIM_threshold} R. Mirzoyan et al, NIM A 387 (1997) 74-78
\bibitem{bib:max_PMT} M. Kn\"{o}tig et al (2011) CTA  ̇Sensors ̇ ICATPP ̇2011


\end{thebibliography}
\end{document}